\documentclass[12pt,fleqn]{article}
\usepackage{amssymb}
\newcommand{\vek}[1]{\mathrm{\bf #1}}

\newcommand{\be}{\begin{equation}}
\newcommand{\ee}{\end{equation}}

\begin{document}
\begin{center}
\section*{Supertranslations to all orders}
Rainer Dick\\[5mm]
Department of Physics and Engineering Physics, 
University of Saskatchewan,\\
116 Science Place, Saskatoon, SK S7N 5E2, Canada\\[10mm]
\end{center}

{\bf Abstract:} 
The transformation laws of the general linear superfield
$V(x,\theta,\overline{\theta})$ and chiral superfields
under $N=1$ supertranslations $\exp[\mathrm{i}(\zeta\cdot Q
+\overline{Q}\cdot\overline{\zeta})]$ are tabulated
to all orders in the parameters $\zeta$, $\overline{\zeta}$.\\[5mm]
{\it Keywords:} Supersymmetry; Supermultiplets; Supersymmetric
transformation laws\\[5mm]
PACS: 11.30.Pb

%%%%%%%%%%%%%%%%%%%%%%%%%%%%%%%%%%%%%%%%%%%%%%%%%%%%%%%
\section{Introduction}\label{sec:intro}
%%%%%%%%%%%%%%%%%%%%%%%%%%%%%%%%%%%%%%%%%%%%%%%%%%%%%%%

Quantum field theories with exact correspondences between bosonic
and fermionic helicity states are not only basic ingredients for
superstring theories, but have dominated both theoretical investigations
and experimental searches for particle physics beyond the current
``Standard Model'' of particle physics for over three decades now.

The minimal version of supersymmetric extensions of the Standard
Model extends the generators $M_{\mu\nu}$, $p_\mu$
of the Poincar\'e group by a set of fermionic generators 
$Q_\alpha$ and $Q_{\dot{\alpha}}$ in the $(\frac{1}{2},0)$
and $(0,\frac{1}{2})$ representations of the proper orthochronous
Lorentz group in four dimensions. It has been recognized early on
that this extension of the Poincar\'e algebra to a super-Poincar\'e
algebra can be represented linearly (and in a reducible, but not
fully reducible manner) on a set comprising 4 complex spin-0 fields, 
4 Weyl spinors, and one complex spin-1 field. This set constitutes the 
so called general linear multiplet or general linear superfield $V$, 
and its irreducible subsets had also been identified.

Indeed, it is entirely sufficient to know the action of the 
supertranslation generators $Q_\alpha$ and $Q_{\dot{\alpha}}$ on the 
components of $V$, or equivalently the action of the supertranslation
$\exp[\mathrm{i}(\zeta\cdot Q+\overline{Q}\cdot\overline{\zeta})]$
to first order in the parameters $\zeta_\alpha$, 
$\overline{\zeta}_{\dot{\alpha}}$, to construct supersymmetric action
principles and the related supercurrents. Therefore the first order 
transformation laws for the components of $V$ can be found in many books 
and review articles on supersymmetry, and with our current understanding 
this is all what is needed to discuss the physical implications of 
supersymmetry. 

However, from a mathematical point of view it seems desirable to have the 
full transformation properties of the general linear multiplet readily 
available for reference. To provide such a reference is the purpose of 
this paper. To make these results also easily accessible for beginners in 
supersymmetry, the super-Poincar\'e algebra and the basic techniques
of superspace calculations, as they pertain to the derivation of the
transformation properties of the component fields, are also 
briefly reviewed\footnote{Every serious student of supersymmetry will
still have to consult Wess and Bagger \cite{WB} to learn the foundations
of the subject.}. 
Therefore the outline of the paper is as follows.

Conventions for spinor representations of the Lorentz group
and the super-Poincar\'e algebra are discussed in section
\ref{sec:sPoinc}.
Superspace is reviewed in section \ref{sec:theta}, and
the full supertranslation properties of the component fields 
of the general linear multiplet are reported in section \ref{sec:results}.

Chiral superfields provide a particular irreducible representation
within the reducible linear multiplet. 
Due to their practical relevance for the supersymmetrization of matter
fields, the resulting full supertranslation properties of the components
of chiral superfields are listed in section \ref{sec:chiral}.

Appendix 1 contains a translation into the conventions of Wess and Bagger
\cite{WB}. The relevant spinor identities are reviewed in appendix 2.

The conventions used here differ from Wess and Bagger 
only with regard to the definition of superderivatives and the
definition of the 2nd order epsilon spinors with lower indices. 
Sections \ref{sec:sPoinc} and \ref{sec:theta} are included to make the 
paper self-contained and easily accessible, and to clarify
conventions. However, the results in sections \ref{sec:results} and 
\ref{sec:chiral} are not affected by the different definitions.
The {\it cognoscenti} should therefore go straight to section
\ref{sec:results} and refer to the earlier sections only if the
need arises. 

There is a further reason besides accessibility, why 
the larger part of this paper is an
introductory review rather than an original research contribution. 
While (to my best knowledge) the full supertranslation laws of the 
component fields have not been published before, the methodology for 
calculations with linear super-multiplets in four dimensions was
developed some 35 years ago by Wess, Zumino, Salam, Strathdee, and 
Ferrara \cite{WZ,SS,FWZ,SS2}. The full 
supertranslation laws should prove useful for the further development of 
supersymmetry and its applications, but tabulating those transformation 
laws primarily closes a gap in the literature.

%%%%%%%%%%%%%%%%%%%%%%%%%%%%%%%%%%%%%%%%%%%%%%%%%%%%%%%
\section{The super-Poincar\'e algebra}\label{sec:sPoinc}
%%%%%%%%%%%%%%%%%%%%%%%%%%%%%%%%%%%%%%%%%%%%%%%%%%%%%%%

We use $\eta_{00}=-1$ for the Minkowski metric and standard
notation $\sigma_\mu{}_{\alpha\dot{\alpha}}$ with
\[
\sigma_0=\left(\sigma_{0\,\alpha\dot{\alpha}}\right)
=\left(\begin{array}{cc}
1 & \,0\\
0 & \,1\\
\end{array}
\right),\quad
\sigma_1=\left(\sigma_{1\,\alpha\dot{\alpha}}\right)
=\left(\begin{array}{cc}
0 & \,1\\
1 & \,0\\
\end{array}
\right),
\]
\[
\sigma_2=\left(\sigma_{2\,\alpha\dot{\alpha}}\right)
=\left(\begin{array}{cc}
0 & \,-\mathrm{i}\\
\mathrm{i} & \,\,\,\,\,0\\
\end{array}
\right),\quad
\sigma_3=\left(\sigma_{3\,\alpha\dot{\alpha}}\right)
=\left(\begin{array}{cc}
1 & \,\,\,\,\,0\\
0 & \,-1\\
\end{array}
\right)
\]
for the Pauli matrices. 

Complex conjugation turns undotted indices into dotted indices 
and vice versa,
\[
(\psi_\alpha)^\ast=\overline{\psi}_{\dot{\alpha}},
\]
and hermiticity of the Pauli matrices implies for the complex
conjugate matrices
\begin{equation}\label{eq:hermit1}
\overline{\sigma}{}^\mu{}_{\dot{\alpha}\alpha}=
\sigma^\mu{}_{\alpha\dot{\alpha}}.
\end{equation}

We pull spinor indices with the two-dimensional epsilon spinors
\begin{equation}\label{eq:epsil}
\epsilon^{12}=\epsilon_{12}=1,\quad
\epsilon^{\dot{1}\dot{2}}=\epsilon_{\dot{1}\dot{2}}=1,
\end{equation}
\begin{equation}\label{eq:pull}
\psi^\alpha=\epsilon^{\alpha\beta}\psi_\beta,\quad
\psi_\beta=\psi^\alpha\epsilon_{\alpha\beta},
\end{equation}
\begin{equation}\label{eq:dotpull}
\overline{\psi}{}^{\dot{\alpha}}=
\epsilon^{\dot{\alpha}\dot{\beta}}\psi_{\dot{\beta}},\quad
\overline{\psi}_{\dot{\beta}}=\overline{\psi}{}^{\dot{\alpha}}
\epsilon_{\dot{\alpha}\dot{\beta}}.
\end{equation}

The equations (\ref{eq:hermit1}) then imply that the conjugate
Pauli matrices with upper spinor indices are
\begin{equation}\label{eq:ovsig}
\overline{\sigma}{}^{\mu\,\dot{\alpha}\alpha}
=\epsilon^{\dot{\alpha}\dot{\beta}}\epsilon^{\alpha\beta}
\sigma^\mu{}_{\beta\dot{\beta}}.
\end{equation}
Numerically we have with the upper index positions for the barred 
matrices and lower index positions for the unbarred matrices,
\[
\overline{\sigma}_0=\sigma_0,\quad\overline{\sigma}_i=-\sigma_i.
\]
Although not formally required, use of upper indices for
barred Pauli matrices and lower indices for unbarred Pauli matrices 
is a useful and very common convention. Relations for Pauli matrices
are meticulously compiled in Ref. \cite{WB}. For convenience, we recall 
those relations which are directly relevant for the derivation of 
supertranslations to all orders,
\begin{equation}\label{eq:sigsig}
(\overline{\sigma}_\mu)^{\dot{\alpha}\alpha}
(\sigma^\mu)_{\beta\dot{\beta}}
=-2\delta^{\dot{\alpha}}{}_{\dot{\beta}}\delta^\alpha{}_\beta,\quad
(\overline{\sigma}_\mu\cdot\sigma^\mu)^{\dot{\alpha}}{}_{\dot{\beta}}
=-4\delta^{\dot{\alpha}}{}_{\dot{\beta}},\quad
(\sigma^\mu\cdot\overline{\sigma}_\mu)_\alpha{}^\beta
=-4\delta_\alpha{}^\beta,
\end{equation}
\begin{equation}\label{eq:sigcliff}
(\sigma_\mu\cdot\overline{\sigma}_\nu
+\sigma_\nu\cdot\overline{\sigma}_\mu)_\alpha{}^\beta
=-2\eta_{\mu\nu}\delta_\alpha{}^\beta,
\quad
(\overline{\sigma}_\mu\cdot\sigma_\nu
+\overline{\sigma}_\nu\cdot\sigma_\mu)^{\dot{\alpha}}{}_{\dot{\beta}}
=-2\eta_{\mu\nu}\delta^{\dot{\alpha}}{}_{\dot{\beta}},
\end{equation}
\begin{equation}\label{eq:tracesig2}
\mathrm{Tr}\!\left(\sigma_\mu\cdot\overline{\sigma}_\nu\right)
=\sigma_{\mu\alpha\dot{\alpha}}
\overline{\sigma}_\nu{}^{\dot{\alpha}\alpha}=-2\eta_{\mu\nu},
\end{equation}
and
\begin{equation}\label{eq:sig3}
\sigma^\lambda\cdot\overline{\sigma}{}^{\mu}\cdot\sigma^\nu
=\eta^{\lambda\nu}\sigma^\mu-\eta^{\lambda\mu}\sigma^\nu
-\eta^{\mu\nu}\sigma^\lambda-\mathrm{i}\epsilon_{0123}
\epsilon^{\lambda\mu\nu\rho}\sigma_\rho.
\end{equation}
The factor $\epsilon_{0123}=\pm 1$
was included to allow for ready use of both conventions for
the four-dimensional $\epsilon$-tensor.

We will briefly recall below that pulling spinor indices with
the 2nd order epsilon spinors is motivated by the fact
that this yields Lorentz invariant spinor products
\begin{equation}\label{eq:prod1}
\psi\cdot\chi\equiv\psi^\alpha\chi_\alpha
=\epsilon^{\alpha\beta}\psi_\beta\chi_\alpha
=-\psi_\alpha\chi^\alpha=\chi^\alpha\psi_\alpha
=\chi\cdot\psi,
\end{equation}
where the anti-commutation property of spinors was used.

Conjugation also implies re-ordering of spinor quantities,
such that conjugation of equation (\ref{eq:prod1}) yields
\begin{equation}\label{eq:prod2}
\overline{\chi}\cdot\overline{\psi}\equiv
\overline{\chi}_{\dot{\alpha}}\overline{\psi}{}^{\dot{\alpha}}
=\overline{\chi}_{\dot{\alpha}}\epsilon^{\dot{\alpha}\dot{\beta}}
\overline{\psi}_{\dot{\beta}}=\overline{\psi}\cdot\overline{\chi}.
\end{equation}

The vector
representation matrices of the Lorentz algebra
\[
(L_{\mu\nu})_\rho{}^\sigma=\eta_{\mu\rho}\eta_\nu{}^\sigma
-\eta_\mu{}^\sigma\eta_{\nu\rho}
\]
appear as structure constants in the Poincar\'e algebra.
The spinor representations of a proper orthochronous Lorentz
transformation
\[
\Lambda=\exp\!\left(\frac{\mathrm{i}}{2}\omega^{\mu\nu}M_{\mu\nu}\right)
=\exp\!\left(\frac{1}{2}\omega^{\mu\nu}L_{\mu\nu}\right)
\]
are given by
\[
U(\Lambda)=\exp\!\left(\frac{\mathrm{i}}{2}
\omega^{\mu\nu}S_{\mu\nu}\right)
\]
and
\[
\tilde{U}(\Lambda)=\exp\!\left(\frac{\mathrm{i}}{2}
\omega^{\mu\nu}\overline{S}_{\mu\nu}\right),
\]
with generators
\[
(S_{\mu\nu})_\alpha{}^\beta=\frac{\mathrm{i}}{4}
(\sigma_\mu\cdot\overline{\sigma}_\nu
-\sigma_\nu\cdot\overline{\sigma}_\mu)_\alpha{}^\beta,
\]
\[
(\overline{S}_{\mu\nu})^{\dot{\alpha}}{}_{\dot{\beta}}
=\frac{\mathrm{i}}{4}(\overline{\sigma}_\mu\cdot\sigma_\nu
-\overline{\sigma}_\nu\cdot\sigma_\mu)^{\dot{\alpha}}{}_{\dot{\beta}}.
\]

The following relations hold,
\begin{equation}\label{eq:TrSS1}
\mathrm{Tr}\!\left(S_{\mu\nu}\cdot S_{\rho\sigma}\right)
=\frac{1}{2}\!\left(\eta_{\mu\rho}\eta_{\nu\sigma}
-\eta_{\mu\sigma}\eta_{\nu\rho}
-\mathrm{i}\epsilon_{0123}\epsilon_{\mu\nu\rho\sigma}\right),
\end{equation}
\begin{equation}\label{eq:TrSS2}
\mathrm{Tr}\!\left(\overline{S}_{\mu\nu}\cdot
\overline{S}_{\rho\sigma}\right)
=\frac{1}{2}\!\left(\eta_{\mu\rho}\eta_{\nu\sigma}
-\eta_{\mu\sigma}\eta_{\nu\rho}
+\mathrm{i}\epsilon_{0123}\epsilon_{\mu\nu\rho\sigma}\right).
\end{equation}
These relations are used in the derivation of supersymmetric
Maxwell or Yang-Mills actions. 

The spinor products (\ref{eq:prod1},\ref{eq:prod2}) are invariant
because the matrices $U(\Lambda)$ and $\tilde{U}(\Lambda)$  are
SL(2,$\mathbb{C}$) matrices,
\begin{equation}\label{eq:sl2c1}
\epsilon^{\alpha\beta}
U(\Lambda)_\alpha{}^\gamma U(\Lambda)_\beta{}^\delta
=\epsilon^{\gamma\delta},\quad
\epsilon_{\dot{\alpha}\dot{\beta}}
\tilde{U}(\Lambda)^{\dot{\alpha}}{}_{\dot{\gamma}}
\tilde{U}(\Lambda)^{\dot{\beta}}{}_{\dot{\delta}}
=\epsilon_{\dot{\gamma}\dot{\delta}}.
\end{equation}
Stated differently: the epsilon spinors are Lorentz invariant.

We can now write down the super-Poincar\'e algebra 
in the form
\[
[M_{\mu\nu},M_{\rho\sigma}]=\mathrm{i}
(L_{\mu\nu})_\rho{}^\lambda M_{\lambda\sigma}+\mathrm{i}
(L_{\mu\nu})_\sigma{}^\lambda M_{\rho\lambda},
\]
\[
[M_{\mu\nu},p_\rho]=\mathrm{i}(L_{\mu\nu})_\rho{}^\sigma p_\sigma,
\]
\[
[M_{\mu\nu},Q_\alpha]=-\frac{\mathrm{i}}{4}
(\sigma_\mu\cdot\overline{\sigma}_\nu
-\sigma_\nu\cdot\overline{\sigma}_\mu)_\alpha{}^\beta Q_\beta,
\]
\begin{equation}\label{eq:MovQ}
[M_{\mu\nu},\overline{Q}{}^{\dot{\alpha}}]=-\frac{\mathrm{i}}{4}
(\overline{\sigma}_\mu\cdot\sigma_\nu
-\overline{\sigma}_\nu\cdot\sigma_\mu)^{\dot{\alpha}}{}_{\dot{\beta}}
\overline{Q}{}^{\dot{\beta}},
\end{equation}
\[
\{Q_\alpha,\overline{Q}_{\dot{\alpha}}\}=2p_{\alpha\dot{\alpha}}
=2p_\mu\sigma^\mu{}_{\alpha\dot{\alpha}},
\]
\[
\{Q_\alpha,Q_\beta\}=0,\quad\{\overline{Q}_{\dot{\alpha}},
\overline{Q}_{\dot{\beta}}\}=0,
\]
\[
[p_\mu,Q_\alpha]=0,\quad
[p_\mu,\overline{Q}_{\dot{\alpha}}]=0.
\]

The SL(2,$\mathbb{C}$) property (\ref{eq:sl2c1})
\begin{eqnarray*}
&&\epsilon^{\gamma\beta}
(\sigma_\mu\cdot\overline{\sigma}_\nu
-\sigma_\nu\cdot\overline{\sigma}_\mu)_\gamma{}^\alpha
+\epsilon^{\alpha\gamma}
(\sigma_\mu\cdot\overline{\sigma}_\nu
-\sigma_\nu\cdot\overline{\sigma}_\mu)_\gamma{}^\beta
\\
&=&(\sigma_\mu\cdot\overline{\sigma}_\nu
-\sigma_\nu\cdot\overline{\sigma}_\mu)^{\alpha\beta}
-(\sigma_\mu\cdot\overline{\sigma}_\nu
-\sigma_\nu\cdot\overline{\sigma}_\mu)^{\beta\alpha}
=0
\end{eqnarray*}
implies that equation (\ref{eq:MovQ}) can also be written as
\[
[M_{\mu\nu},\overline{Q}_{\dot{\alpha}}]
=\frac{\mathrm{i}}{4}\overline{Q}_{\dot{\beta}}
(\overline{\sigma}_\mu\cdot\sigma_\nu
-\overline{\sigma}_\nu\cdot\sigma_\mu)^{\dot{\beta}}{}_{\dot{\alpha}}.
\]

The super-Poincar\'e algebra satisfies all the pertinent
super-Jacobi identities as a consequence of the representation
properties of the vector and spinor representations of the
Lorentz algebra. The particular super-Jacobi identity
\[
[M_{\mu\nu},\{Q_\alpha,\overline{Q}_{\dot{\alpha}}\}]
=\{[M_{\mu\nu},Q_\alpha],\overline{Q}_{\dot{\alpha}}\}
+\{[M_{\mu\nu},\overline{Q}_{\dot{\alpha}}],Q_\alpha\}
\]
holds as a consequence of the fact that the Pauli matrices have the same 
form in every inertial frame,
\begin{eqnarray}\nonumber
(L_{\mu\nu})^\lambda{}_\kappa
\sigma^\kappa{}_{\alpha\dot{\alpha}}&=&
\eta_\mu{}^\lambda\sigma_{\nu\alpha\dot{\alpha}}
-\eta_\nu{}^\lambda\sigma_{\mu\alpha\dot{\alpha}}
\\ \label{invsig}
&=&\frac{1}{4}
(\sigma_\mu\cdot\overline{\sigma}_\nu
-\sigma_\nu\cdot\overline{\sigma}_\mu)_\alpha{}^\beta
\sigma^\lambda{}_{\beta\dot{\alpha}}
-\sigma^\lambda{}_{\alpha\dot{\beta}}
\frac{1}{4}
(\overline{\sigma}_\mu\cdot\sigma_\nu
-\overline{\sigma}_\nu\cdot\sigma_\mu)^{\dot{\beta}}{}_{\dot{\alpha}}.
\end{eqnarray}
This can be verified from equations (\ref{eq:sigcliff}) by commuting
the $\sigma^\lambda$ matrices into the middle positions in the products
on the right hand side. It can also be verified as a direct consequence
of equation (\ref{eq:sig3}).

%%%%%%%%%%%%%%%%%%%%%%%%%%%%%%%%%%%%%%%%%%%%%%%%%%%%%%%
\section{$N=1$ Superspace}\label{sec:theta}
%%%%%%%%%%%%%%%%%%%%%%%%%%%%%%%%%%%%%%%%%%%%%%%%%%%%%%%

The Poincar\'e algebra is realized on spacetime coordinates $x^\mu$
through derivative operators
\begin{equation}\label{eq:dpoinc}
M_{\mu\nu}=-\mathrm{i}(x_\mu\partial_\nu-x_\nu\partial_\mu),\quad
p_\mu=-\mathrm{i}\partial_\mu.
\end{equation}

In a nutshell, superspace is based on the observation that this
construction can be extended to the super-Poincar\'e algebra
by supplementing Minkowski spacetime with fermionic coordinates
coordinates $\theta^\alpha$ and $\overline{\theta}{}^{\dot{\alpha}}$
and corresponding fermionic derivatives
\begin{equation}\label{eq:defd}
\partial_\alpha\theta^\beta=\delta_\alpha{}^\beta,
\quad
\overline{\partial}_{\dot{\alpha}}\overline{\theta}{}^{\dot{\beta}}
=\delta_{\dot{\alpha}}{}^{\dot{\beta}}.
\end{equation}

The super-Poincar\'e algebra is then realized on the superspace
coordinates $(x^\mu,\theta^\alpha,\overline{\theta}{}^{\dot{\alpha}})$ 
by amending the representations (\ref{eq:dpoinc}) of the bosonic operators
with the realizations
\begin{eqnarray}\label{eq:Qder1}
Q_\alpha&=&-\mathrm{i}\partial_\alpha
-\sigma^\mu{}_{\alpha\dot{\alpha}}\overline{\theta}{}^{\dot{\alpha}}
\partial_\mu,
\\ \label{eq:Qder2}
\overline{Q}_{\dot{\alpha}}&=&
\mathrm{i}\overline{\partial}_{\dot{\alpha}}
+\theta^\alpha\sigma^\mu{}_{\alpha\dot{\alpha}}\partial_\mu,
\end{eqnarray}
for the fermionic operators, and complementing the Lorentz generators 
to include the action on $Q_\alpha$ and $\overline{Q}_{\dot{\alpha}}$,
\[
M_{\mu\nu}=-\mathrm{i}\left(x_\mu\partial_\nu-x_\nu\partial_\mu\right)
+\frac{\mathrm{i}}{4}\theta^\alpha
(\sigma_\mu\cdot\overline{\sigma}_\nu
-\sigma_\nu\cdot\overline{\sigma}_\mu)_\alpha{}^\beta\partial_\beta
-\frac{\mathrm{i}}{4}
(\overline{\sigma}_\mu\cdot\sigma_\nu
-\overline{\sigma}_\nu\cdot\sigma_\mu)^{\dot{\beta}}{}_{\dot{\alpha}}
\overline{\theta}{}^{\dot{\alpha}}\overline{\partial}_{\dot{\beta}}.
\]

Superfields $V(x,\theta,\overline{\theta})$ are complex valued
functions on superspace.
The identities (\ref{eq:TT1}-\ref{eq:ident3}) imply that 
the expansion of every superfield with respect to the fermionic
variables $(\theta^\alpha,\overline{\theta}{}^{\dot{\alpha}})$ 
can be written in terms of four scalars
$\phi(x)$, $M(x)$, $N(x)$, $D(x)$, four Weyl fermions
$\psi(x)$, $\chi(x)$, $\kappa(x)$, $\lambda(x)$, 
and a vector field $A_\mu(x)$,
\begin{eqnarray}\nonumber
V\!\left(x,\theta,\overline{\theta}\right)&=&
\phi(x)+\theta\cdot\psi(x)+\overline{\chi}(x)\cdot\overline{\theta}
+\theta\cdot\sigma^\mu\cdot\overline{\theta}A_\mu(x)
+\theta^2 M(x)+\overline{\theta}{}^2 N(x)
\\ \label{eq:V1}
&&+\overline{\theta}{}^2\theta\cdot\kappa(x)
+\theta^2\overline{\lambda}(x)\cdot\overline{\theta}
+\theta^2\overline{\theta}{}^2 D(x).
\end{eqnarray}

The commutation relations
\[
\mathrm{i}[\zeta\cdot Q,\theta^\alpha]
=\zeta^\alpha,\quad
\mathrm{i}[\overline{Q}\cdot
\overline{\zeta},\overline{\theta}{}^{\dot{\alpha}}]
=\overline{\zeta}{}^{\dot{\alpha}},
\]
\[
\mathrm{i}[\zeta\cdot Q
+\overline{Q}\cdot\overline{\zeta},x^\mu]
=-\mathrm{i}\left(\zeta\cdot\sigma^\mu\cdot\overline{\theta}
-\theta\cdot\sigma^\mu\cdot\overline{\zeta}\right)
\]
imply for unitary supertranslations
\begin{eqnarray*}
|x',\theta',\overline{\theta}{}'\rangle
&=&\exp\!\left[\mathrm{i}\!\left(\zeta\cdot Q
+\overline{Q}\cdot\overline{\zeta}\right)\right]
|x,\theta,\overline{\theta}\rangle
\\
&=&
|x+\mathrm{i}(\zeta\cdot\sigma\cdot\overline{\theta}
-\theta\cdot\sigma\cdot\overline{\zeta}),
\theta-\zeta,\overline{\theta}-\overline{\zeta}\rangle,
\end{eqnarray*}
and therefore
\begin{eqnarray}\nonumber
V'\!\left(x,\theta,\overline{\theta}\right)&=&
\langle x,\theta,\overline{\theta}|
\exp\!\left[\mathrm{i}\!\left(
\zeta\cdot Q+\overline{Q}\cdot\overline{\zeta}\right)\right]
|V\rangle
\\ \label{eq:V2}
&=&V\!\left(x-\mathrm{i}(\zeta\cdot\sigma\cdot\overline{\theta}
-\theta\cdot\sigma\cdot\overline{\zeta}),
\theta+\zeta,\overline{\theta}+\overline{\zeta}\right).
\end{eqnarray}

We can calculate the transformation properties of the
component fields by comparing
\begin{eqnarray*}
V'\!\left(x,\theta,\overline{\theta}\right)&=&
\phi'(x)+\theta\cdot\psi'(x)
+\overline{\chi}'(x)\cdot\overline{\theta}
+\theta\cdot\sigma^\mu\cdot\overline{\theta}A'_\mu(x)
+\theta^2 M'(x)+\overline{\theta}{}^2 N'(x)
\\
&&+\overline{\theta}{}^2\theta\cdot\kappa'(x)
+\theta^2\overline{\lambda}'(x)\cdot\overline{\theta}
+\theta^2\overline{\theta}{}^2 D'(x)
\end{eqnarray*}
with the expansion of the right hand side of equation (\ref{eq:V2}) 
with respect to the fermionic variables $\theta^\alpha$ 
and $\overline{\theta}{}^{\dot{\alpha}}$.

%%%%%%%%%%%%%%%%%%%%%%%%%%%%%%%%%%%%%%%%%%%%%%%%%%%%%%%
\section{Supertranslations of the general linear multiplet}
\label{sec:results}
%%%%%%%%%%%%%%%%%%%%%%%%%%%%%%%%%%%%%%%%%%%%%%%%%%%%%%%

Supertranslations shift the argument $x$ of component fields to
\[
X^\mu=x^\mu-\mathrm{i}\!\left(\zeta\cdot\sigma^\mu\cdot\overline{\theta}
-\theta\cdot\sigma^\mu\cdot\overline{\zeta}\right).
\]
We can calulate the transformation properties of the components
of the supermultiplet $V$ to all orders in the transformation
parameters $\zeta$, $\overline{\zeta}$, by expanding the right
hand side
\begin{eqnarray*}
V\!\left(X,\theta+\zeta,\overline{\theta}+\overline{\zeta}\right)
&=&\phi\!\left(x-\mathrm{i}(\zeta\cdot\sigma\cdot\overline{\theta}
-\theta\cdot\sigma\cdot\overline{\zeta})\right)
\\
&&+\left(\theta+\zeta\right)\cdot
\psi\!\left(x-\mathrm{i}(\zeta\cdot\sigma\cdot\overline{\theta}
-\theta\cdot\sigma\cdot\overline{\zeta})\right)
+\ldots
\end{eqnarray*}
of equation (\ref{eq:V2}) in all orders in $\theta$ and
$\overline{\theta}$. 

The first step requires the expansion of component fields
with respect to the coordinate shift 
$\Delta x^\mu=\mathrm{i}\!\left(\zeta\cdot\sigma^\mu\cdot
\overline{\theta}-\theta\cdot\sigma^\mu\cdot\overline{\zeta}\right)$,
e.g.
\begin{eqnarray}\nonumber
\phi(X)&=&\phi(x)-\mathrm{i}
\left(\zeta\cdot\sigma^\mu\cdot\overline{\theta}
-\theta\cdot\sigma^\mu\cdot\overline{\zeta}\right)\partial_\mu\phi(x)
\\ \nonumber
&&-\frac{1}{2}\left(\zeta\cdot\sigma^\mu\cdot\overline{\theta}
-\theta\cdot\sigma^\mu\cdot\overline{\zeta}\right)
\left(\zeta\cdot\sigma^\nu\cdot\overline{\theta}
-\theta\cdot\sigma^\nu\cdot\overline{\zeta}\right)
\partial_\mu\partial_\nu\phi(x)
\\ \nonumber
&&+\frac{\mathrm{i}}{6}\left(\zeta\cdot\sigma^\lambda
\cdot\overline{\theta}
-\theta\cdot\sigma^\lambda\cdot\overline{\zeta}\right)
\left(\zeta\cdot\sigma^\mu\cdot\overline{\theta}
-\theta\cdot\sigma^\mu\cdot\overline{\zeta}\right)
\left(\zeta\cdot\sigma^\nu\cdot\overline{\theta}
-\theta\cdot\sigma^\nu\cdot\overline{\zeta}\right)
\\ \nonumber
&&\times\partial_\lambda\partial_\mu\partial_\nu\phi(x)
+\frac{1}{24}\left(\zeta\cdot\sigma^\kappa
\cdot\overline{\theta}
-\theta\cdot\sigma^\kappa\cdot\overline{\zeta}\right)
\left(\zeta\cdot\sigma^\lambda
\cdot\overline{\theta}
-\theta\cdot\sigma^\lambda\cdot\overline{\zeta}\right)
\\ \nonumber
&&\times\left(\zeta\cdot\sigma^\mu\cdot\overline{\theta}
-\theta\cdot\sigma^\mu\cdot\overline{\zeta}\right)
\left(\zeta\cdot\sigma^\nu\cdot\overline{\theta}
-\theta\cdot\sigma^\nu\cdot\overline{\zeta}\right)
\partial_\kappa\partial_\lambda\partial_\mu\partial_\nu\phi(x)
\\ \nonumber
&=&\phi(x)-\mathrm{i}
\left(\zeta\cdot\sigma^\mu\cdot\overline{\theta}
-\theta\cdot\sigma^\mu\cdot\overline{\zeta}\right)\partial_\mu\phi(x)
\\ \nonumber
&&+\frac{1}{4}\left(\zeta^2\overline{\theta}{}^2
+\theta^2\overline{\zeta}{}^2
+2\zeta\cdot\sigma_\mu\cdot\overline{\zeta}
\theta\cdot\sigma^\mu\cdot\overline{\theta}\right)\partial^2\phi(x)
\\ \nonumber
&&-\frac{1}{2}\left(\zeta\cdot\sigma^\nu
\cdot\overline{\zeta}\theta\cdot\sigma^\mu\cdot\overline{\theta}
+\zeta\cdot\sigma^\mu\cdot\overline{\zeta}
\theta\cdot\sigma^\nu\cdot\overline{\theta}\right)
\partial_\mu\partial_\nu\phi(x)
\\ \nonumber
&&-\frac{\mathrm{i}}{4}\theta^2\overline{\zeta}{}^2
\zeta\cdot\sigma^\mu\cdot\overline{\theta}\partial_\mu\partial^2\phi(x)
+\frac{\mathrm{i}}{4}\zeta^2\overline{\theta}{}^2
\theta\cdot\sigma^\mu\cdot\overline{\zeta}\partial_\mu\partial^2\phi(x)
\\ \label{eq:expphi}
&&+\frac{1}{16}\theta^2\overline{\theta}{}^2\zeta^2
\overline{\zeta}{}^2\partial^2\partial^2\phi(x),
\end{eqnarray} 
and corresponding expansions %%(\ref{eq:exppsi1}-\ref{eq:expD9})
for combinations of the other eight component fields with various
factors, which are different in each case due to the presence of
fermionic variables in the extra factors. Altogether, this includes
35 more relations, e.g. 
\begin{eqnarray}\nonumber
\theta\cdot\psi(X)&=&\theta\cdot\psi(x)
-\frac{\mathrm{i}}{2}
\theta\cdot\sigma^\mu\cdot\overline{\theta}
\zeta\cdot\sigma^\nu\cdot\overline{\sigma}_\mu\cdot
\partial_\nu\psi(x)
+\frac{\mathrm{i}}{2}\theta^2\overline{\zeta}
\cdot\overline{\sigma}{}^\mu\cdot\partial_\mu\psi(x)
\\ \nonumber
&&+\frac{1}{4}\zeta^2\overline{\theta}{}^2\theta^\alpha
\partial^2\psi_\alpha(x)
+\frac{1}{4}\theta^2\zeta\cdot\sigma_\mu\cdot\overline{\zeta}
\overline{\theta}\cdot\overline{\sigma}{}^\mu\cdot
\partial^2\psi(x)
\\ \nonumber
&&-\frac{1}{4}\theta^2\left(
\zeta\cdot\sigma^\mu\cdot\overline{\zeta}\overline{\theta}
\cdot\overline{\sigma}{}^\nu
+\zeta\cdot\sigma^\nu\cdot\overline{\zeta}\overline{\theta}
\cdot\overline{\sigma}{}^\mu
\right)\cdot\partial_\mu\partial_\nu\psi(x)
\\ \label{eq:exppsi1}
&&-\frac{\mathrm{i}}{8}\theta^2\overline{\theta}{}^2
\zeta^2\partial_\mu\partial^2\psi(x)\cdot\sigma^\mu\cdot
\overline{\zeta}.
\end{eqnarray}

Substitution of all the expansions in terms of standard words in the
Grassmann variables into equation (\ref{eq:V2}) yields the full 
supertranslation properties of the component fields, which are reported
in equations (\ref{eq:phi'}-\ref{eq:D'}). The transformation equations 
of the component fields are organized by contributions from the nine 
component fields $\phi(x)$, $\psi(x)$, $\overline{\chi}(x)$, $A_\mu(x)$, 
$M(x)$, $N(x)$, $\kappa(x)$, $\overline{\lambda}(x)$, and $D(x)$,
instead of organization by expansion in the supertranslation parameters
$\zeta$ and $\overline{\zeta}$. In this way, supertranslations act on
the component fields like matrices which have Grassmann valued 
differential operators as entries. The reader can easily re-organize
the transformation equations in terms of supertranslation parameters.

The transformation equations are
\begin{eqnarray}\nonumber
\phi'(x)&=&\phi(x)+\zeta\cdot\psi(x)
+\overline{\chi}(x)\cdot\overline{\zeta}
+\zeta\cdot\sigma^\mu\cdot\overline{\zeta} A_\mu(x)
+\zeta^2 M(x)
\\ \label{eq:phi'}
&&+\overline{\zeta}{}^2N(x)
+\overline{\zeta}{}^2\zeta\cdot\kappa(x)
+\zeta^2\overline{\zeta}\cdot\overline{\lambda}(x)
+\zeta^2\overline{\zeta}{}^2 D(x),
\end{eqnarray}

\begin{eqnarray}\nonumber
\psi'(x)&=&\mathrm{i}\sigma^\mu
\cdot\overline{\zeta}\partial_\mu\phi(x)+\psi(x)
+\frac{\mathrm{i}}{2}\sigma^\mu
\cdot\overline{\sigma}_\nu\cdot\partial_\mu\psi(x)
\zeta\cdot\sigma^\nu\cdot\overline{\zeta}
-\frac{\mathrm{i}}{2}\overline{\zeta}{}^2\sigma^\mu\cdot\partial_\mu
\overline{\chi}
\\ \nonumber
&&+\sigma^\mu\cdot\overline{\zeta}A_\mu(x)+\frac{\mathrm{i}}{2}
\sigma^\nu\cdot\overline{\sigma}{}^\mu\cdot\zeta\overline{\zeta}{}^2
\partial_\nu A_\mu(x)+2\zeta M(x)+\mathrm{i}\zeta^2\sigma^\mu
\cdot\overline{\zeta}\partial_\mu M(x)
\\ \label{eq:psi'}
&&+\kappa(x)\overline{\zeta}{}^2-\sigma_\mu
\cdot\overline{\lambda}(x)\zeta\cdot\sigma^\mu\cdot
\overline{\zeta}-\frac{\mathrm{i}}{2}
\zeta^2\overline{\zeta}{}^2\sigma^\mu
\cdot\partial_\mu\overline{\lambda}(x)
+2\zeta\overline{\zeta}{}^2 D(x),
\end{eqnarray}

\begin{eqnarray}\nonumber
\overline{\chi}'(x)&=&-\mathrm{i}\partial_\mu\phi(x)
\zeta\cdot\sigma^\mu
+\frac{\mathrm{i}}{2}\zeta^2\partial_\mu\psi(x)\cdot\sigma^\mu
+\overline{\chi}(x)-\frac{\mathrm{i}}{2}\zeta\cdot\sigma^\nu\cdot
\overline{\zeta}\partial_\mu\overline{\chi}(x)\cdot
\overline{\sigma}_\nu\cdot\sigma^\mu
\\ \nonumber
&&+A_\mu(x)\zeta\cdot\sigma^\mu-\frac{\mathrm{i}}{2}\zeta^2
\partial_\nu A_\mu(x)\overline{\zeta}\cdot\overline{\sigma}{}^\mu
\cdot\sigma^\nu+2\overline{\zeta}N(x)-\mathrm{i}
\overline{\zeta}{}^2\partial_\mu N(x)\zeta\cdot\sigma^\mu
\\ \label{eq:chi'}
&&-\zeta\cdot\sigma^\mu\cdot\overline{\zeta}\kappa(x)\cdot\sigma_\mu
+\frac{\mathrm{i}}{2}\zeta^2\overline{\zeta}{}^2\partial_\mu\kappa(x)
\cdot\sigma^\mu+\zeta^2\overline{\lambda}(x)+2\zeta^2\overline{\zeta}D(x),
\end{eqnarray}

\begin{eqnarray}\nonumber
A'_\mu(x)&=&\frac{1}{2}\zeta\cdot\sigma_\mu\cdot\overline{\zeta}
\partial^2\phi(x)
-\zeta\cdot\sigma^\nu\cdot\overline{\zeta}
\partial_\mu\partial_\nu\phi(x)
\\ \nonumber
&&-\frac{\mathrm{i}}{2}\zeta\cdot\sigma^\nu\cdot
\overline{\sigma}_\mu\cdot\partial_\nu\psi(x)
-\frac{1}{4}\zeta^2\partial^2\psi(x)
\cdot\sigma_\mu\cdot\overline{\zeta}
+\frac{1}{2}\zeta^2\partial_\mu\partial_\nu\psi(x)
\cdot\sigma^\nu\cdot\overline{\zeta}
\\ \nonumber
&&+\frac{\mathrm{i}}{2}\partial_\nu\overline{\chi}(x)\cdot
\overline{\sigma}_\mu\cdot\sigma^\nu\cdot\overline{\zeta}
-\frac{1}{4}\overline{\zeta}{}^2\zeta\cdot\sigma_\mu\cdot
\partial^2\overline{\chi}(x)
+\frac{1}{2}\overline{\zeta}{}^2\zeta\cdot\sigma^\nu\cdot
\partial_\mu\partial_\nu\overline{\chi}(x)
\\ \nonumber
&&+A_\mu(x)+\frac{\mathrm{i}}{2}\zeta\cdot\sigma^\lambda
\cdot\overline{\sigma}_\mu\cdot\sigma^\nu\cdot\overline{\zeta}
\left(\partial_\nu A_\lambda(x)-\partial_\lambda A_\nu(x)\right)
\\ \nonumber
&&-\frac{1}{4}\zeta^2\overline{\zeta}{}^2\left(
\partial^2 A_\mu(x)-2\partial_\mu\partial_\nu A^\nu(x)\right)
+\mathrm{i}\zeta^2\partial_\mu M(x)
-\mathrm{i}\overline{\zeta}{}^2\partial_\mu N(x)
\\ \nonumber
&&-\kappa(x)\cdot\sigma_\mu\cdot\overline{\zeta}
-\mathrm{i}\overline{\zeta}{}^2\zeta\cdot\partial_\mu\kappa(x)
-\frac{\mathrm{i}}{2}\overline{\zeta}{}^2\zeta\cdot\sigma^\nu
\cdot\overline{\sigma}_\mu\cdot\partial_\nu\kappa(x)
\\ \nonumber
&&-\zeta\cdot\sigma_\mu\cdot\overline{\lambda}(x)
+\mathrm{i}\zeta^2\overline{\zeta}\cdot\partial_\mu
\overline{\lambda}(x)
+\frac{\mathrm{i}}{2}\zeta^2\partial_\nu\overline{\lambda}(x)
\cdot\overline{\sigma}_\mu\cdot\sigma^\nu\cdot\overline{\zeta}
\\ \label{eq:Amu'}
&&-2\zeta\cdot\sigma_\mu\cdot\overline{\zeta} D(x).
\end{eqnarray}
 For conversion of the last equation
into standard ``words'' in the Grassmann variables
$\zeta$ and $\overline{\zeta}$, note that from equation
(\ref{eq:sig3})
\[
\frac{\mathrm{i}}{2}\zeta\cdot\sigma^\lambda
\cdot\overline{\sigma}_\mu\cdot\sigma^\nu\cdot\overline{\zeta}
 F_{\nu\lambda}(x)=\frac{1}{2}\epsilon_{0123}
\epsilon^\lambda{}_\mu{}^\nu{}_\rho\zeta\cdot\sigma^\rho\cdot
\overline{\zeta}F_{\nu\lambda}(x)
=\epsilon_{0123}\tilde{F}_{\mu\rho}(x)\zeta\cdot\sigma^\rho\cdot
\overline{\zeta}.
\]

The remaining transformation equations are
\begin{eqnarray}\nonumber
M'(x)&=&\frac{1}{4}\overline{\zeta}{}^2\partial^2\phi(x)
-\frac{\mathrm{i}}{2}\partial_\mu\psi(x)\cdot\sigma^\mu
\cdot\overline{\zeta}
+\frac{1}{4}\overline{\zeta}{}^2\zeta\cdot\partial^2\psi(x)
-\frac{\mathrm{i}}{2}\overline{\zeta}{}^2\partial_\mu A^\mu(x)
\\ \nonumber
&&+M(x)-\mathrm{i}\zeta\cdot\sigma^\mu\cdot\overline{\zeta}
\partial_\mu M(x)+\frac{1}{4}\zeta^2\overline{\zeta}{}^2\partial^2 M(x)
\\ \label{eq:M'}
&&+\overline{\zeta}\cdot\overline{\lambda}(x)
+\frac{\mathrm{i}}{2}\overline{\zeta}{}^2\zeta\cdot\sigma^\mu\cdot
\partial_\mu\overline{\lambda}(x)+\overline{\zeta}{}^2 D(x),
\end{eqnarray}

\begin{eqnarray}\nonumber
N'(x)&=&\frac{1}{4}\zeta^2\partial^2\phi(x)
+\frac{\mathrm{i}}{2}\zeta\cdot\sigma^\mu\cdot\partial_\mu
\overline{\chi}(x)
+\frac{1}{4}\zeta^2\overline{\zeta}\cdot\partial^2
\overline{\chi}(x)
+\frac{\mathrm{i}}{2}\zeta^2\partial_\mu A^\mu(x)
\\ \nonumber
&&+N(x)+\mathrm{i}\zeta\cdot\sigma^\mu\cdot
\overline{\zeta}\partial_\mu N(x)
+\frac{1}{4}\zeta^2\overline{\zeta}{}^2\partial^2 N(x)
\\ \label{eq:N'}
&&+\zeta\cdot\kappa(x)-\frac{\mathrm{i}}{2}\zeta^2
\partial_\mu\kappa(x)\cdot\sigma^\mu\cdot\overline{\zeta}
+\zeta^2 D(x),
\end{eqnarray}

\begin{eqnarray}\nonumber
\kappa'(x)&=&\frac{\mathrm{i}}{4}\zeta^2
\sigma^\mu\cdot\overline{\zeta}\partial_\mu\partial^2\phi(x)
+\frac{1}{4}\zeta^2\partial^2\psi(x)
-\frac{1}{4}(\zeta\cdot\sigma_\mu\cdot\overline{\zeta})
\sigma^\mu\cdot\partial^2\overline{\chi}(x)
\\ \nonumber
&&+\frac{1}{2}(\zeta\cdot\sigma^\mu\cdot\overline{\zeta})
\sigma^\nu\cdot\partial_\mu\partial_\nu\overline{\chi}(x)
-\frac{\mathrm{i}}{8}\zeta^2\overline{\zeta}{}^2
\sigma^\mu\cdot
\partial_\mu\partial^2\overline{\chi}(x)
\\ \nonumber
&&-\frac{\mathrm{i}}{2}\sigma^\mu\cdot\overline{\sigma}{}^\nu
\cdot\zeta\partial_\nu A_\mu(x)
+\frac{1}{4}\zeta^2\sigma^\mu\cdot\overline{\zeta}
\partial^2 A_\mu(x)-\frac{1}{2}\zeta^2
\sigma^\mu\cdot\overline{\zeta}\partial_\mu\partial_\nu A^\nu(x)
\\ \nonumber
&&+\mathrm{i}\sigma^\mu\cdot\overline{\zeta}
\partial_\mu N(x)+\frac{1}{2}\zeta\overline{\zeta}{}^2
\partial^2 N(x)
\\ \nonumber
&&+\kappa(x)+\frac{\mathrm{i}}{2}
(\zeta\cdot\sigma^\nu\cdot\overline{\zeta})\sigma^\mu\cdot
\overline{\sigma}_\nu\cdot\partial_\mu\kappa(x)
+\mathrm{i}
(\zeta\cdot\sigma^\mu\cdot\overline{\zeta})\partial_\mu\kappa(x)
\\ \label{eq:kappa'}
&&-\frac{\mathrm{i}}{2}\zeta^2\sigma^\mu\cdot\partial_\mu
\overline{\lambda}(x)
+2\zeta D(x),
\end{eqnarray}

\begin{eqnarray}\nonumber
\overline{\lambda}'(x)&=&
-\frac{\mathrm{i}}{4}\overline{\zeta}{}^2
\partial_\mu\partial^2\phi(x)\zeta\cdot\sigma^\mu
-\frac{1}{4}(\zeta\cdot\sigma_\mu\cdot
\overline{\zeta})\partial^2\psi(x)\cdot\sigma^\mu
\\ \nonumber
&&+\frac{1}{2}(\zeta\cdot\sigma^\mu\cdot\overline{\zeta})
\partial_\mu\partial_\nu\psi(x)\cdot\sigma^\nu
+\frac{\mathrm{i}}{8}\zeta^2\overline{\zeta}{}^2
\partial_\mu\partial^2\psi(x)\cdot\sigma^\mu
\\ \nonumber
&&+\frac{1}{4}\overline{\zeta}{}^2\partial^2\overline{\chi}(x)
+\frac{\mathrm{i}}{2}\partial_\nu A_\mu(x)
\overline{\zeta}\cdot\overline{\sigma}{}^\nu\cdot\sigma^\mu
+\frac{1}{4}\overline{\zeta}{}^2\partial^2 A_\mu(x)
\zeta\cdot\sigma^\mu
\\ \nonumber
&&-\frac{1}{2}\overline{\zeta}{}^2\zeta\cdot
\sigma^\mu\partial_\mu\partial_\nu A^\nu(x)
-\mathrm{i}\partial_\mu M(x)\zeta\cdot\sigma^\mu
+\frac{1}{2}\zeta^2\overline{\zeta}\partial^2 M(x)
\\ \nonumber
&&+\frac{\mathrm{i}}{2}\overline{\zeta}{}^2\partial_\mu\kappa(x)
\cdot\sigma^\mu
+\overline{\lambda}(x)
-\frac{\mathrm{i}}{2}(\zeta\cdot\sigma^\nu\cdot\overline{\zeta})
\partial_\mu\overline{\lambda}(x)
\cdot\overline{\sigma}_\nu\cdot\sigma^\mu
\\ \label{eq:lambda'}
&&-\mathrm{i}(\zeta\cdot\sigma^\mu\cdot\overline{\zeta})
\partial_\mu\overline{\lambda}(x)
+2\overline{\zeta}D(x),
\end{eqnarray}

\begin{eqnarray}\nonumber 
D'(x)&=&\frac{1}{16}\zeta^2 
\overline{\zeta}{}^2\partial^2\partial^2\phi(x)
-\frac{\mathrm{i}}{8}
\zeta^2\partial_\mu\partial^2\psi(x)\cdot\sigma^\mu\cdot
\overline{\zeta}+\frac{\mathrm{i}}{8}
\overline{\zeta}{}^2\zeta\cdot\sigma^\mu\cdot
\partial_\mu\partial^2\overline{\chi}(x)
\\ \nonumber
&&-\frac{1}{4}\zeta\cdot\sigma^\mu\cdot\overline{\zeta}
\left(\partial^2 A_\mu(x)-2\partial_\mu\partial_\nu A^\nu(x)\right)
\\ \nonumber
&&+\frac{1}{4}\zeta^2\partial^2 M(x)
+\frac{1}{4}\overline{\zeta}{}^2\partial^2 N(x)
-\frac{\mathrm{i}}{2}
\partial_\mu\kappa(x)\cdot\sigma^\mu\cdot\overline{\zeta}
\\ \label{eq:D'}
&&+\frac{\mathrm{i}}{2}
\zeta\cdot\sigma^\mu\cdot\partial_\mu\overline{\lambda}(x)
+D(x).
\end{eqnarray}

These transformation laws are compatible with the reality
constraints $V(x,\theta,\overline{\theta})
=V^+(x,\theta,\overline{\theta})$ which define the vector
multiplet,
\[
\phi(x)=\phi^+(x),\quad \psi(x)=\chi(x),\quad A_\mu(x)=A^+_\mu(x),
\quad M(x)=N^+(x),\quad\kappa(x)=\lambda(x),
\]
\[
D=D^+(x).
\]

%%%%%%%%%%%%%%%%%%%%%%%%%%%%%%%%%%%%%%%%%%%%%%%%%%%%%%%
\section{Supertranslations of the chiral multiplet}\label{sec:chiral}
%%%%%%%%%%%%%%%%%%%%%%%%%%%%%%%%%%%%%%%%%%%%%%%%%%%%%%%

Besides the superderivatives (\ref{eq:Qder1},\ref{eq:Qder2})
one can also define the supercovariant derivatives \cite{FWZ,SS2}
\begin{equation}\label{eq:Dder}
D_\alpha=-\mathrm{i}\partial_\alpha
+\sigma^\mu{}_{\alpha\dot{\alpha}}\overline{\theta}{}^{\dot{\alpha}}
\partial_\mu,\quad
\overline{D}_{\dot{\alpha}}=
\mathrm{i}\overline{\partial}_{\dot{\alpha}}
-\theta^\alpha\sigma^\mu{}_{\alpha\dot{\alpha}}\partial_\mu,
\end{equation}
such that
\[
\{D_\alpha,D_\beta\}=0,\quad
\{\overline{D}_{\dot{\alpha}},\overline{D}_{\dot{\beta}}\}=0,
\quad
\{D_\alpha,\overline{D}_{\dot{\alpha}}\}
=2\mathrm{i}\sigma^\mu{}_{\alpha\dot{\alpha}}\partial_\mu,
\]
and
\[
\{Q_\alpha,\overline{D}_{\dot{\alpha}}\}=0,\quad
\{D_\alpha,\overline{Q}_{\dot{\alpha}}\}=0,\quad
\{Q_\alpha,D_\beta\}=0\quad
\{\overline{Q}_{\dot{\alpha}},\overline{D}_{\dot{\beta}}\}=0.
\]
The condition for chiral superfields
\[
\left(\mathrm{i}\overline{\partial}_{\dot{\alpha}}
-\theta^\alpha\sigma^\mu{}_{\alpha\dot{\alpha}}\partial_\mu\right)
\Phi(x,\theta,\overline{\theta})=0
\]
is therefore invariant under supertranslations.

The basic solutions  
\[
\overline{D}_{\dot{\alpha}}\theta^\alpha=0,\quad
\overline{D}_{\dot{\alpha}}\left(x^\mu+\mathrm{i}
\theta^\alpha\sigma^\mu{}_{\alpha\dot{\beta}}
\overline{\theta}{}^{\dot{\beta}}\right)=0
\]
imply \cite{WB}
\begin{eqnarray*}
\Phi(x,\theta,\overline{\theta})&=&
\Phi(x+\mathrm{i}\theta\cdot\sigma\cdot
\overline{\theta},\theta)
\\
&=&\phi(x+\mathrm{i}\theta\cdot\sigma\cdot
\overline{\theta})
+\theta\cdot\psi(x+\mathrm{i}\theta\cdot\sigma\cdot
\overline{\theta})
+\theta^2 F(x+\mathrm{i}\theta\cdot\sigma\cdot
\overline{\theta}).
\end{eqnarray*}
The equation 
\[
-\theta^\alpha\sigma^\mu{}_{\alpha\dot{\alpha}}
\overline{\theta}{}^{\dot{\alpha}}
\theta^\beta\sigma^\nu{}_{\beta\dot{\beta}}
\overline{\theta}{}^{\dot{\beta}}\partial_\mu\partial_\nu
=\frac{1}{2}\theta^2\overline{\theta}{}^2
\eta^{\mu\nu}\partial_\mu\partial_\nu
\]
yields 
\begin{eqnarray}\nonumber
\Phi(x,\theta,\overline{\theta})&=&
\phi(x)+\mathrm{i}\theta\cdot\sigma^\mu\cdot
\overline{\theta}\partial_\mu\phi(x)
+\frac{1}{4}\theta^2\overline{\theta}{}^2\partial^2\phi(x)
+\theta\cdot\psi(x)
\\ \label{eq:chiral1}
&&-\frac{\mathrm{i}}{2}\theta^2
\partial_\mu\psi(x)\cdot\sigma^\mu\cdot\overline{\theta}
+\theta^2 F(x).
\end{eqnarray}

The chiral superfield corresponds to the following substitutions 
in the general superfield $V$,
\[
\phi\to\phi,\quad\psi\to\psi,\quad\overline{\chi}\to 0,\quad
A_\mu\to\mathrm{i}\partial_\mu\phi,\quad
M\to F,\quad N\to 0,
\]
\[
\kappa\to 0,\quad
\overline{\lambda}\to
-\frac{\mathrm{i}}{2}\partial_\mu\psi\cdot\sigma^\mu,
\quad D\to\frac{1}{4}\partial^2\phi.
\]
It is clear from the construction, but can also be checked explicitly
that these constraints are compatible with the transformation
properties (\ref{eq:phi'}-\ref{eq:D'}) of the full linear multiplet.

We find the following supertranslations of
the components of the chiral multiplet,
\begin{eqnarray}\nonumber
\phi'(x)&=&\phi(x)+\mathrm{i}\zeta\cdot\sigma^\mu\cdot
\overline{\zeta}\partial_\mu\phi(x)+\frac{1}{4}
\zeta^2\overline{\zeta}{}^2\partial^2\phi(x)
+\zeta\cdot\psi(x)-\frac{\mathrm{i}}{2}\zeta^2
\partial_\mu\psi(x)\cdot\sigma^\mu\cdot\overline{\zeta}
\\ \label{eq:chirtrans1}
&&+\zeta^2 F(x),
\end{eqnarray}
\begin{eqnarray}\nonumber
\psi'(x)&=&2\mathrm{i}\sigma^\mu\cdot\overline{\zeta}
\partial_\mu\phi(x)+\zeta\overline{\zeta}{}^2
\partial^2\phi(x)+\psi(x)
\\ \nonumber
&&+\frac{\mathrm{i}}{2}(\zeta\cdot
\sigma_\nu\cdot\overline{\zeta})\left(
\sigma^\mu\cdot\overline{\sigma}{}^\nu
-\sigma^\nu\cdot\overline{\sigma}{}^\mu\right)\cdot\partial_\mu\psi(x)
-\frac{1}{4}\zeta^2\overline{\zeta}{}^2\partial^2\psi(x)
\\ \label{eq:chirtrans2}
&&+2\zeta F(x)+\mathrm{i}\zeta^2\sigma^\mu\cdot\overline{\zeta}
\partial_\mu F(x),
\end{eqnarray}
\begin{eqnarray}\nonumber
F'(x)&=&\overline{\zeta}{}^2\partial^2\phi(x)
-\mathrm{i}\partial_\mu\psi(x)\cdot\sigma^\mu\cdot
\overline{\zeta}+\frac{1}{2}\overline{\zeta}{}^2\zeta\cdot
\partial^2\psi(x)
\\ \label{eq:chirtrans3}
&&+F(x)-\mathrm{i}\zeta\cdot\sigma^\mu\cdot
\overline{\zeta}\partial_\mu F(x)+\frac{1}{4}
\zeta^2\overline{\zeta}{}^2\partial^2 F(x).
\end{eqnarray}

Please note that this presentation does not involve the usual 
rescaling of  the spinor component in the chiral superfield,
\[
\psi(x)\to\sqrt{2}\psi(x),
\]
which is required for canonically normalized kinetic terms.

%%%%%%%%%%%%%%%%%%%%%%%%%%%%%%%%%%%%%%%%%%%%%%%%%%%%%%%
\section{Conclusions}\label{sec:conc}
%%%%%%%%%%%%%%%%%%%%%%%%%%%%%%%%%%%%%%%%%%%%%%%%%%%%%%%

The supertranslation properties of the component fields of
a general linear supermultiplet and of a chiral multiplet were 
reported to all orders in the translation parameters $\zeta$
and $\overline{\zeta}$ in equations (\ref{eq:phi'}-\ref{eq:D'})
and equations (\ref{eq:chirtrans1}-\ref{eq:chirtrans3}),
respectively. On the one hand, one can think of these results
as explicit parametrizations of orbits of supertranslations
in the space of component fields of a supersymmetric theory.
On the other hand, one can consider the transformed fields
again as superfields in variables $(x,\zeta,\overline{\zeta})$,
because e.g.
\begin{eqnarray*}
\psi'_\alpha(x)&=&\psi'_\alpha(x,\zeta,\overline{\zeta})
=\frac{\partial}{\partial\theta^\alpha}
V\!\left(x-\mathrm{i}(\zeta\cdot\sigma\cdot\overline{\theta}
-\theta\cdot\sigma\cdot\overline{\zeta}),\theta+\zeta,
\overline{\theta}+\overline{\zeta}\right)\Big|_{\theta=0,\overline{\theta}=0}
\\
&=&\left(\mathrm{i}\sigma^\mu{}_{\alpha\dot{\alpha}}
\overline{\zeta}{}^{\dot{\alpha}}\partial_\mu
+\frac{\partial}{\partial\zeta^\alpha}\right)
V\!\left(x,\zeta,\overline{\zeta}\right)
=\mathrm{i}D_\alpha^{(\zeta)}V\!\left(x,\zeta,\overline{\zeta}\right),
\end{eqnarray*}
and higher order derivatives with respect to the $\theta$
and $\overline{\theta}$ variables at $\theta=0$, $\overline{\theta}=0$
can also be expressed as supercovariant derivatives
with respect to $\zeta$ and $\overline{\zeta}$. From this point of view, 
equations (\ref{eq:phi'}-\ref{eq:D'}) and 
(\ref{eq:chirtrans1}-\ref{eq:chirtrans3}) tell us explicitly
how the fields $\phi(x)$, $\psi(x)$, $A_\mu(x)$, $M(x)$, $\kappa(x)$
and $D(x)$ induce superfields $\phi'(x,\zeta,\overline{\zeta})$,
$\psi'(x,\zeta,\overline{\zeta})$, $A'_\mu(x,\zeta,\overline{\zeta})$, 
$M'(x,\zeta,\overline{\zeta})$, $\kappa'(x,\zeta,\overline{\zeta})$
and $D'(x,\zeta,\overline{\zeta})$.

%%%%%%%%%%%%%%%%%%%%%%%%%%%%%%%%%%%%%%%%%%%%%%%%%%%%%%%
\section*{Appendix 1}
%%%%%%%%%%%%%%%%%%%%%%%%%%%%%%%%%%%%%%%%%%%%%%%%%%%%%%%

Our superspace realizations (\ref{eq:Qder1},\ref{eq:Qder2})
and (\ref{eq:Dder}) are related to the
realizations in Wess and Bagger \cite{WB} according to
\[
Q_\alpha=-\mathrm{i}Q_\alpha^{(WB)},\quad
\overline{Q}_{\dot{\alpha}}=-\mathrm{i}\overline{Q}{}_{\dot{\alpha}}^{(WB)},
\]
\[
D_\alpha=-\mathrm{i}D_\alpha^{(WB)},\quad
\overline{D}_{\dot{\alpha}}=-\mathrm{i}\overline{D}_{\dot{\alpha}}^{(WB)}.
\]
With these conventions, the mode expansions for chiral superfields agree,
and the generators for supertranslations are also the same
(cf. equations (4.11,4.12) in \cite{WB}),
\begin{eqnarray*}
\mathrm{i}\!\left(\zeta\cdot Q+\overline{Q}\cdot\overline{\zeta}\right)
&=&\zeta^\alpha\partial_\alpha-\overline{\partial}_{\dot{\alpha}}
\overline{\zeta}{}^{\dot{\alpha}}-\mathrm{i}\!\left(\zeta\cdot\sigma^\mu
\cdot\overline{\theta}-\theta\cdot\sigma^\mu
\cdot\overline{\zeta}\right)\partial_\mu
\\
&=&\zeta\cdot Q^{(WB)}+\overline{Q}{}^{(WB)}\cdot\overline{\zeta},
\end{eqnarray*}
i.e. our results for supertranslations to all orders also apply
in the conventions of Wess and Bagger.

Note that Wess and Bagger use an operator representation
of the super-Poincar\'e algebra with the same signature of the
Minkowski metric but $p_\mu=\mathrm{i}\partial_\mu$. 
This comes from the familiar sign difference between field theoretic
and quantum mechanical operator realizations of Noether charges. If
$\phi(x)=\langle x|\phi\rangle$ is a field operator, the
momentum operators
\[
P_\mu=\int\!d^3\vek{x}\left(\eta_\mu{}^0\mathcal{L}
-\partial_\mu\phi\cdot
\frac{\partial\mathcal{L}}{\partial(\partial_0\phi)}\right)
\]
satisfy
\[
[P_\mu,\phi(x)]=\mathrm{i}\partial_\mu\phi(x)
\]
and generate translations according to
\begin{eqnarray*}
\phi'(x)&=&
\exp(-\mathrm{i}\epsilon\cdot P)\phi(x)\exp(\mathrm{i}\epsilon\cdot P)
=\phi(x+\epsilon)=\langle x+\epsilon|\phi\rangle
\\
&=&\langle x|\exp(\mathrm{i}\epsilon\cdot p)|\phi\rangle
=\langle x|\phi'\rangle.
\end{eqnarray*}
Similar equations hold for Lorentz and gauge charges.
Another way to look at the sign difference is through 
Jacobi identities. If the generators $X_a$ satisfy the
Lie algebra
\[
[X_a,X_b]=\mathrm{i}C_{ab}{}^c X_c,
\]
then the adjoint matric representation is given by
\[
(X_a)_b{}^c=-\mathrm{i}C_{ab}{}^c.
\]

%%%%%%%%%%%%%%%%%%%%%%%%%%%%%%%%%%%%%%%%%%%%%%%%%%%%%%%
\section*{Appendix 2}
%%%%%%%%%%%%%%%%%%%%%%%%%%%%%%%%%%%%%%%%%%%%%%%%%%%%%%%

There are several useful identities for products of spinors
which are used in the determination of the general linear multiplet
and its transformations laws. 
The following identities are a direct consequence of the
anti-symmetry properties of spinors and the definitions
(\ref{eq:prod1}), (\ref{eq:prod2}) of spinor products\footnote{Please
note that $\theta^2$ in the first two equations of (\ref{eq:TT1})
denotes the component $\theta^{\alpha=2}$ of the spinor $\theta$, but in the
last of these equations it is $\theta^2=\theta\cdot\theta$. In every
equation in supersymmetry it is clear from the context what $\theta^2$
means. However, readers without experience in supersymmetric calculations
can take comfort in the fact that in the previous sections of this paper 
$\theta^2$ always refers to $\theta^2=\theta\cdot\theta$.},
\begin{equation}\label{eq:TT1}
\theta^\alpha\theta^\beta=\epsilon^{\alpha\beta}\theta^1\theta^2
=\frac{1}{2}\epsilon^{\alpha\beta}
\left(-\theta^1\theta_1+\theta_2\theta^2\right)
=-\frac{1}{2}\epsilon^{\alpha\beta}\theta^\gamma\theta_\gamma
=-\frac{1}{2}\epsilon^{\alpha\beta}\theta^2,
\end{equation}
\begin{equation}\label{eq:TT2}
\overline{\theta}{}^{\dot{\alpha}}\overline{\theta}{}^{\dot{\beta}}
=\epsilon^{\dot{\alpha}\dot{\beta}}
\overline{\theta}{}^{\dot{1}}\overline{\theta}{}^{\dot{2}}
=\frac{1}{2}\epsilon^{\dot{\alpha}\dot{\beta}}\overline{\theta}{}^2,
\end{equation}
\begin{equation}\label{eq:TT3}
\theta_\alpha\theta^\beta=-\theta^\beta\theta_\alpha
=-\frac{1}{2}\delta_\alpha{}^\beta\theta^2,\quad
\overline{\theta}_{\dot{\alpha}}\overline{\theta}{}^{\dot{\beta}}
=-\overline{\theta}{}^{\dot{\beta}}\overline{\theta}_{\dot{\alpha}}
=\frac{1}{2}\delta_{\dot{\alpha}}{}^{\dot{\beta}}\overline{\theta}{}^2.
\end{equation}

The following relations also use the properties 
(\ref{eq:ovsig}-\ref{eq:tracesig2}) of Pauli matrices,
\begin{equation}\label{eq:prop1}
\psi\cdot\sigma^\mu\cdot\overline{\chi}=\psi^\alpha
\sigma^\mu{}_{\alpha\dot{\alpha}}\overline{\chi}{}^{\dot{\alpha}}
=\epsilon^{\alpha\beta}\epsilon^{\dot{\alpha}\dot{\beta}}
\sigma^\mu{}_{\alpha\dot{\alpha}}\psi_\beta\overline{\chi}_{\dot{\beta}}
=-\overline{\chi}\cdot\overline{\sigma}^\mu\cdot\psi,
\end{equation}
\begin{equation}\label{eq:prop2}
\psi\cdot\sigma^\mu\cdot\overline{\sigma}{}^\nu\cdot\chi
=\epsilon^{\alpha\beta}\psi_\beta
\sigma^\mu{}_{\alpha\dot{\alpha}}
\epsilon^{\dot{\alpha}\dot{\beta}}
\sigma^\nu{}_{\gamma\dot{\beta}}\epsilon^{\delta\gamma}\chi_\delta
=\chi^\gamma\sigma^\nu{}_{\gamma\dot{\beta}}
\overline{\sigma}{}^{\mu \dot{\beta}\beta}\psi_\beta
=\chi\cdot\sigma^\nu\cdot\overline{\sigma}{}^{\mu}\cdot\psi,
\end{equation}
\begin{equation}\label{eq:prop3}
\overline{\theta}{}^{\dot{\alpha}}\psi^\alpha
=-\frac{1}{2}\overline{\sigma}_\lambda{}^{\dot{\alpha}\alpha}
\sigma^\lambda{}_{\gamma\dot{\gamma}}
\overline{\theta}{}^{\dot{\gamma}}\psi^\gamma
=\frac{1}{2}\overline{\sigma}_\lambda{}^{\dot{\alpha}\alpha}
\psi\cdot\sigma^\lambda\cdot\overline{\theta},
\end{equation}
\begin{equation}\label{eq:ident3}
\zeta^\alpha\sigma^\mu{}_{\alpha\dot{\alpha}}
\overline{\theta}{}^{\dot{\alpha}}\zeta^\beta
\sigma^\nu{}_{\beta\dot{\beta}}\overline{\theta}{}^{\dot{\beta}}
=\frac{1}{4}\epsilon^{\alpha\beta}\epsilon^{\dot{\alpha}\dot{\beta}}
\sigma^\mu{}_{\alpha\dot{\alpha}}\sigma^\nu{}_{\beta\dot{\beta}}
\zeta^2\overline{\theta}{}^2=-\frac{1}{2}\eta^{\mu\nu}
\zeta^2\overline{\theta}{}^2.
\end{equation}

\vspace*{5mm}

{\bf Acknowledgment:} This work was supported by NSERC Canada.

\end{document}